\documentclass[aps,10pt]{revtex4}

\begin{document}
\thispagestyle{empty}
\title{Space with spinor structure
and analytical properties of the solutions of Klein-Fock and
Schr\"{o}dinger equations in cylindric parabolic  coordinates }
\author{V.M. Red'kov\\ Institute of Physics, National  Academy of Sciences of Belarus}

\begin{abstract}
Possible  quantum mechanical  corollaries of  changing the
vectorial  geometri\-cal model of the  physical space, extending
it twice,  in order to describe its spinor structure (in other
terminology and emphasis it is known as the Hopf's bundle) are
investigated. The extending procedure is realized in cylindrical
parabolic coordinates: $G(t,u,v,z) \Longrightarrow
\tilde{G}(t,u,v,z)$. It is done through expansion twice as  much
of the domain $G$  so that instead of the  half plane $(u,v>0)$
now the entire plane  $(u,v)$ should  be used accompanied  with
new identification rules over the  boundary points. In the
Cartesian picture this procedure there corresponds to taking the
two-sheet  surface $(x',y') \oplus (x'',y'')$ in place of the
one-sheet surface  $(x,y)$. Solutions of the Klein-Fock and
Schr\"{o}dinger equations $ \Psi _{\epsilon, p, \;a}  =
e^{i\epsilon t} e^{ipz} U_{a}(u) V_{a}(v) $ are constructed in
terms of parabolic cylinder functions, $a$ is a separating
constant. Given quantum numbers  $\epsilon, p, a$ four types of
solutions are possible: $\Psi_{++}, \Psi_{--}; $ $\;
\Psi_{+-},\Psi_{-+}$. The  first  two $\Psi_{++}, \Psi_{--}$
provide us with single-valued functions of the vectorial space
points, whereas last two $\Psi_{+-},\Psi_{-+}$ have
discon\-tinuities in the frame of vectorial space and therefore
they must be  rejected in this   model. All four types of
functions  are continuous ones  being regarded in  the spinor
space. Explicit form of a 2-order differential operator
diagonalized on the  constructed  wave functions with  the
eigenvalue $a$: $\hat{A} \; \Psi_{\epsilon, p, } = +a \;
\Psi_{\epsilon, p, a}$ is found both in  $(t,x,y,z)$ and
$(t,u,v,z)$ representation. It is shown that solutions $\Psi_{++},
\Psi_{--}$, $ \; \Psi_{+-},\Psi_{-+}$ all are  the eigen-functions
of two discrete spinor operators $\hat{\delta}$ and  $\hat{\pi}$:
$ \hat{\delta} \; (u,v) =(-u,-v)\; , \;\;  \hat{\pi} (u,v)
=(u,-v)\; , \; \hat{\delta} \; (x,y) =(x,y) \;, \;  \hat{\pi}
(x,y) =(x,-y)\;. $ Two other classifications of the wave functions
over discrete quantum numbers are given. It is established that
all  solutions
 $\Psi_{++}, \Psi_{--}, \; \Psi_{+-},\Psi_{-+}$ are orthogonal to each other provided that
integration is  done over extended domain parameterizing the
spinor space. Simple selection rules for matric elements of the
vector and  spinor coordinates, $(x,y)$ and $(u,v)$, respectively,
are  derived. Selection rules for  $(u,v)$ are substantially
different in vector and spinor spaces. In the supplement some
relationships describing primary geometric objects,
 spatial spinor  $\xi$  and  $\eta$, as functions of  cylindrical parabolic coordinates,
 are given.

\end{abstract}

\maketitle


%\newcommand{\N}{N\raise.7ex\hbox{\underline{$\circ $}}$\;$}

%V.M. Red'kov \\}
%redkov@dragon.bas-net.by

%1
\section{Introduction}

\hspace{5mm} In the literature, there exist    [1-31] three
terminological different approaches though  close in their
intrinsic essence. There are a space-time spinor structure
(see the book by  by Penrose and Rindler [29]  as a modern embodiment of the
old idea [1-4] to use spinor groups  instead of orthogonal); the
Hooph bundle [5]; Kustaaheimo-Stifel bundle [6,7].

Differences between three mentioned formalisms consist mainly in
conceptual accents (see for more detail [32]). In the Hopf's
technique it  is suggested to use in all parts only complex spinors
$\xi$ and conjugated  $\xi^{*}$ instead of real-valued vector
(tensor) quantities. In the Kustaanheimo-Stifel approach we are to
use four real-valued coordinates, form which
Cartesian coordinates  $(x,y,z)$ can be formed
up by means of definite bilinear functions.
These four variables by Kustaanheimo-Stifel  are real and imaginary
parts of two spinor components. The known spinor invariant
$(\xi' \xi^{'*}+ \xi^{2}\xi^{'2})$ becomes the  sum of four squared real quantities,
so that we can associate spinor technique with geometry of  the  Riemann
 space $S_{3}$ of constant positive curvature.

In essence, the Kustaanheimo-Stifel's approach is other elaboration of the same
Hopf's technique based on complex spinors  $\xi$ and
$\xi^{*}$,  in terms of four real-valued variables.
In so doing,  we are able to hide in the formalism  the presence  of the
non-analytical operation of complex conjugation. Spinor space structure,
formalism developed in the  present work, also exploits possibilities given by spinors
to construct 3-vectors, however the emphasis is taken to doubling
the set of spatial points so that we get an
extended space model that is called a space with spinor structure [32-38].
In such an extended space, in place of $2\pi$-rotation, only $4\pi$-rotation transfers
the space into itself.

The procedure itself of doubling the manyfold can be realized easier when for parameterizing
the space some curvilinear coordinate system is used instead of the Cartesian coordinates.
In such context, spherical and parabolic coordinates were considered in [37]. In the present
paper, the use of cylindrical parabolic coordinates is studied as applied for description
of spinor space structure. Now we study  analytical properties of
Schr\"{o}dinger and Klein-Fock  the wave  solutions
depending on vector and spinor space models.
It is demonstrated explicitly that transition to an extended space model
(with spinor structure) lead us to augmenting the number of basis wave functions
of a quantum-mechanical scalar particle.
Also, some possible manifestations of the extended space structure in matrix elements
of physical quantities are discussed.

%2
\section{ Parabolic cylindrical coordinates }

\hspace{5mm} These coordinates in the vector 3-space model
are introduced by relations
\begin{eqnarray}
x = {u^{2} - v^{2} \over 2}  \; , \qquad y = u \; v\; , \qquad z=
z  \; .
\label{2.1}
\\
v^{2} = - x + \sqrt{x^{2} + y^{2}} \; , \qquad u^{2} = + x +
\sqrt{x^{2} + y^{2}} \; .
\label{2.2b}
\end{eqnarray}

To cover all points of  the vector space  $(x,y,z)$ it
suffices any one from  the following four solutions:
\begin{eqnarray}
v = + \sqrt{ - x + \sqrt{x^{2} + y^{2}} } \; , \qquad u = \pm
\sqrt{
 + x + \sqrt{x^{2} + y^{2}} }\; ,
\label{2.3}
\\
v = - \sqrt{ - x + \sqrt{x^{2} + y^{2}} } \; , \qquad u = \pm
\sqrt{
 + x + \sqrt{x^{2} + y^{2}} }\; ,
\nonumber
\\
v = \pm \sqrt{ - x + \sqrt{x^{2} + y^{2}} } \; , \qquad u = +
\sqrt{
 + x + \sqrt{x^{2} + y^{2}} }\; ,
\nonumber
\\
v = \pm \sqrt{ - x + \sqrt{x^{2} + y^{2}} } \; , \qquad u = -
\sqrt{
 + x + \sqrt{x^{2} + y^{2}} }\; .
\end{eqnarray}

\noindent For definiteness, let us  use the first variant from (\ref{2.3}):
\begin{eqnarray}
v = + \sqrt{ - x + \sqrt{x^{2} + y^{2}} } \; , \qquad u = \pm
\sqrt{
 + x + \sqrt{x^{2} + y^{2}} }\; .
\nonumber
\end{eqnarray}

\unitlength=0.4mm
\begin{picture}(100,50)(-120,0)
\special{em:linewidth 0.4pt} \linethickness{0.6pt}

\put(-50,0){\vector(+1,0){100}}  \put(70,-5){$u$}
\put(0,-50){\vector(0,+1){100}}  \put(-10, +45){$v$}

\put(+1,+1){\line(+1,+1){40}} \put(+21,+1){\line(+1,+1){40}}
\put(+21,+1){\line(+1,+1){40}} \put(-19,+1){\line(+1,+1){40}}
\put(-38,+1){\line(+1,+1){40}} \put(-58,+1){\line(+1,+1){40}}
\put(+41,+1){\line(+1,+1){40}}

\end{picture}

\vspace{20mm}

\begin{center}
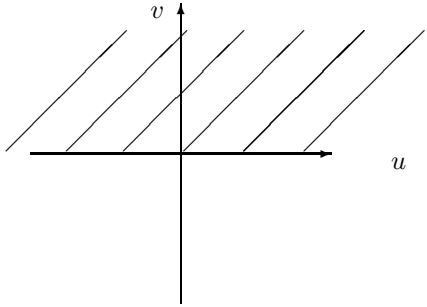
{\bf Fig 1. The domain $G(u,v)$ to  parameterize the vector model  }
\end{center}

Correspondence between the points $(x,y)$  and  $(u,v)$
can be illustrated by the formulas and schemes:
\begin{eqnarray}
u = k \; \cos  \phi \;  , \qquad  v = k \;\sin \phi \; ,
\qquad \phi \in [\; 0 , \; \pi\; ] \; ;
\nonumber
\\
x = (k^{2}/2)\; cos 2\phi \; , \qquad y = (k^{2}/2) \; \sin 2\phi \; ,
\qquad
 2\phi  \in  [ 0 , \; 2 \pi ]\;
\end{eqnarray}

\vspace{5mm}

\unitlength=0.4mm
\begin{picture}(100,50)(-80,0)
\special{em:linewidth 0.4pt} \linethickness{0.6pt}

\put(-50,0){\vector(+1,0){100}}  \put(65,-5){$x$}
\put(0,-50){\vector(0,+1){100}}  \put(-10, +45){$y$}

\put(-30,+25){$B_{1}$}  \put(-30,-25){$B_{2}$}
\put(+25,+25){$A_{1}$}  \put(+25,-25){$A_{2}$}

\put(+100,0){\vector(+1,0){100}}  \put(215,-5){$u$}
\put(150,0){\vector(0,+1){50}}  \put(140, +45){$v$}
\put(150,0){\line(+1,+1){40}}  \put(150,0){\line(-1,+1){40}}

\put(180,+10){$A_{1}$}   \put(110,+10){$A_{2}$}
\put(155,+20){$B_{1}$}   \put(135,+20){$B_{2}$}

\put(-1.5,+20){$*$}          \put(+165,+15){$*$}
\put(0,-20){\circle*{3}}      \put(+135,+15){\circle*{3}}
\put(+30,0){\circle{6}}       \put(+180,0){\oval(7,7)[t]}
\put(+120,0){\oval(7,7)[t]}

\end{picture}

\vspace{15mm}

\begin{center}
{\bf Fig 2. The mapping  $G(x,y) \Longrightarrow G(u,v)$}
\end{center}

%\vspace{10mm}

In the following, when turning to the case of spinor space, we will see the complete symmetry
 between coordinates $u$ и $v$:
namely, they are referred to Cartesian coordinates of  the extended model
$(x,y,z)\oplus (x,y,z)$ through the formulas (compare with the previous)
\begin{eqnarray}
v = \pm \sqrt{ - x + \sqrt{x^{2} + y^{2}} } \; , \qquad u = \pm
\sqrt{
 + x + \sqrt{x^{2} + y^{2}} }\; .
\label{2.4}
\end{eqnarray}

\noindent
the latter can be illustrated by the Fig 3:

\vspace{5mm}

\unitlength=0.4mm
\begin{picture}(100,50)(-120,0)
\special{em:linewidth 0.4pt} \linethickness{0.6pt}

\put(-50,0){\vector(+1,0){100}}  \put(70,-5){$u$}
\put(0,-50){\vector(0,+1){100}}  \put(-10, +45){$v$}

\put(+1,+1){\line(+1,+1){40}} \put(+21,+1){\line(+1,+1){40}}
\put(+21,+1){\line(+1,+1){40}} \put(-19,+1){\line(+1,+1){40}}
\put(-38,+1){\line(+1,+1){40}} \put(-58,+1){\line(+1,+1){40}}
\put(+41,+1){\line(+1,+1){40}}

\put(+1,-1){\line(+1,-1){40}} \put(+21,-1){\line(+1,-1){40}}
\put(+21,-1){\line(+1,-1){40}} \put(-19,-1){\line(+1,-1){40}}
\put(-38,-1){\line(+1,-1){40}} \put(-58,-1){\line(+1,-1){40}}
\put(+41,-1){\line(+1,-1){40}}

\end{picture}

\vspace{18mm}

\begin{center}
{\bf Fig 3. $\tilde{G}(u,v)$ to cover spinor space }
\end{center}

\noindent The metric of 3-space in parabolic cylindrical coordinates is
\begin{eqnarray}
dl^{2} = dx^{2} + dy^{2} + dz^{2} = (u^{2}+ v^{2}) (du^{2}+  d
v^{2}) +  dz^{2} \; ;
%\label{1.5a}
\nonumber
\end{eqnarray}

\noindent correspondingly, the Minkowsky metric looks as
\begin{eqnarray}
dS^{2} = (dx^{0})^{2} - dl^{2} = c^{2} dt^{2} - (u^{2}+ v^{2})
(du^{2}+  d v^{2}) -  dz^{2} \; .
%\eqno(1.5b)
\nonumber
\end{eqnarray}

%3
\section{ Solutions of the Klein-Fock  equation  and functions of parabolic cylinder}

\hspace{5mm} Let us consider the Klein-Fock equation specified for cylindric parabolic
coordinates:
\begin{eqnarray}
\left [ \;    -  {1 \over c^{2}} {\partial ^{2} \over \partial t^{2}}
 + { \partial ^{2} \over \partial z^{2}}  + {1
\over u^{2} + v^{2}} \; ( {\partial^{2} \over \partial u^{2} } +
 {\partial^{2} \over \partial v^{2} } )
 -{m^{2}c^{2} \over  \hbar^{2}} \right ] \; \Psi = 0 \; ,
 \label{3.1a}
\end{eqnarray}

\noindent After  separating the variables  $(t,z)$  from $(u,v)$ by the
substitution
$
\Psi (t,u,  v, \phi ) = e^{-i\epsilon t / \hbar} \; e^{ipz/\hbar } \; U(u) \;
V(v) \;
$
one gets
\begin{eqnarray}
\left [ \;  {1 \over U } \; {d^{2} U \over  d u^{2}} +
( {\epsilon^{2} \over \hbar^{2} c^{2}}  - { m^{2} c^{2} \over   \hbar^{2}}  - {p^{2}\over \hbar^{2}} ) \; u^{2} \;
 \right]  +
\left [ \;  {1 \over V } \; {d^{2} V \over  d v^{2}} +
({\epsilon^{2}\over \hbar^{2} c^{2}}  -
{m^{2}c^{2} \over  \hbar^{2}}  - {p^{2}\over \hbar^{2}} )    \; v^{2} \;  \right ]  = 0 \; .
 \label{3.1b}
 \end{eqnarray}

\noindent
In the following, the notation is used
\begin{eqnarray}
\lambda^{2} = ( {\epsilon^{2} \over \hbar^{2} c^{2}}  - { m^{2}
c^{2} \over   \hbar^{2}}  - {p^{2}\over \hbar^{2}} ) \; , \qquad [
\lambda ] = {1 \over \mbox{meter}} \; .
\nonumber
\end{eqnarray}

\noindent
Introducing two separation constants $a$  and   $b$   ($a+b=0$), from (\ref{3.1b})
 we can derive two separate equations in variables $u$ and $v$ respectively:
\begin{eqnarray}
{d^{2} U \over  d u^{2}} + (\;  \lambda^{2}  \; u^{2} -
  a \; ) \; U =0 \; , \qquad
{d^{2} V \over  d v^{2}} + ( \; \lambda^{2} \; v^{2}  - b
 \; ) \; V =0 \; .
\label{3.2}
\end{eqnarray}

\noindent
Canonical form of  differential equation of  parabolic cylinder  (type 2, [39]) is
\begin{eqnarray}
{d^{2} F \over d \xi^{2} } \; + \;( \; {\xi^{2} \over 4} - \alpha \; ) \; F
= 0 \; .
 \label{3.3}
\end{eqnarray}

\noindent
Transition in equations ( (\ref{3.2}) to  the canonical form is reached
through the use of dimensionless variables)
\begin{eqnarray}
\sqrt{2\lambda } \; u  \;\; \rightarrow  \;\; u \;   , \qquad
{a \over 2\lambda } \;\; \rightarrow  \;\; a \; ,\;\;
\sqrt{2 \lambda } \; v  \;\;  \rightarrow  \;\; v  \; , \qquad
{b \over 2 \lambda }  \;\; \rightarrow  \;\;  b \;  .
\label{3.4}
\end{eqnarray}

\noindent  So that equations (\ref{3.2}) will take the form:
\begin{eqnarray}
{d^{2} U  \over d u^{2} } \; +  \; ( \;  {u^{2} \over 4} - a \; )
\; U = 0 \; ,
\qquad
{d^{2} V  \over d v^{2} } \; + \; (\;  {v^{2} \over 4} - b \; ) \;
V = 0 \; .
\label{3.5}
\end{eqnarray}

As known, solutions of equation (\ref{3.3}) can be  found as a series:
\begin{eqnarray}
F(\xi) = c_{0} \; +\; c_{1} \;\xi \; +\; c_{2}\; \xi^{2} \;+ \; \sum
_{k=1,2,...}
 \; c_{2k+1} \; \xi^{2k+1} \; +\;
\sum _{k=1,2,...} \; c_{2k+2} \;\xi^{2k+2} \; ;
\label{3.6a}
\end{eqnarray}

\noindent in  (\ref{3.6a}) the terms  of even and odd powers of $\xi$ are
distinguished. After substituting  (\ref{3.6a}) into (\ref{3.3}) we get:
\begin{eqnarray}
\left [ \; c_{2} 2 \;+\; \sum _{k=1,2,...} \; c_{2k+1} (2k+1) (2k)
\; \xi^{2k-1}\; +\; \sum _{k=1,2,...} \; c_{2k+2} (2k+2) (2k+1)
\;\xi^{2k} \; \right ]\; +
\nonumber
\\
{1\over 4} \; \left [ \; c_{0}\; \xi^{2} + c_{1}\; \xi^{3} + c_{2}
\;\xi^{4} \;+\; \sum _{k=1,2,...} \; c_{2k+1} \;\xi^{2k+3}\;+ \sum
_{k=1,2,...} \; c_{2k+2}\; \xi^{2k+4} \; \right ]\; -
\nonumber
\\
- \alpha \; \left [ c_{0} \;+\; c_{1}\; \xi + c_{2} \;\xi^{2} \;+ \;\sum
_{k=1,2,...} \; c_{2k+1} \;\xi^{2k+1} \; +\; \sum _{k=1,2,...} \;
c_{2k+2} \; \xi^{2k+2} \; \right ] = 0 \; ,
\end{eqnarray}

\noindent or separating terms of even and odd powers
\begin{eqnarray}
\left [ \; c_{2} 2 \; + \; \sum _{k=1,2,...} \; c_{2k+2} (2k+2)
(2k+1)\; \xi^{2k} \;+\; {1\over 4} \; c_{0}\;  \xi^{2} \;+ \; {1\over
4} c_{2} \; \xi^{4} \; + \right.
\nonumber
\\
\left. + \; {1 \over 4}\; \sum _{k=1,2,...} \; c_{2k+2} \;
\xi^{2k+4} \; - \; \alpha  \; c_{0} \; -\; \alpha \; c_{2} \; \xi^{2}\; - \alpha \;
\sum _{k=1,2,...} \; c_{2k+2} \; \xi^{2k+2} \; \right
]_{\mbox{even}}\;+
\nonumber
\\
+ \left [ \; \sum _{k=1,2,...} \; c_{2k+1} (2k+1) (2k) \; \xi^{2k-1}
\; +\; {1 \over 4}\;  c_{1} \;\xi^{3}\; + {1 \over 4}\;  \sum
_{k=1,2,...} \; c_{2k+1}\; \xi^{2k+3}\; - \right.
\nonumber
\\
\left. -  \;\alpha  c_{1}\; \xi \; -\; \alpha  \sum _{k=1,2,...} \; c_{2k+1}
\; \xi^{2k+1} \; \right ] _{\mbox{odd}}\; = 0 \;  ,
\end{eqnarray}

\noindent and further
\begin{eqnarray}
\left [ \xi^{0} (2    c_{2} - \alpha c_{0} ) +\xi^{2} (c_{4} \; 4 \times 3  +
{c_{0} \over 4} - \alpha \; c_{2}) \; + \xi^{4}\; ( c_{6}\; 6\times 5 +
{c_{2} \over 4} - \alpha  c_{4} ) + \right.
\nonumber
\\
\left. + \sum _{k=3,4,...}  c_{2k+2} (2k+2) (2k+1)\; \xi^{2k} + {1
\over 4} \sum _{k=1,2,...}  c_{2k+2} \; \xi^{2k+4} - \alpha  \sum
_{k=2,3,...} c_{2k+2} \; \xi^{2k+2} \; \right ]_{\mbox{even}}\;+
\nonumber
\\
\left [ \;  \xi (c_{3} 3 \times 2  - \alpha \; c_{1}) + \xi^{3} (c_{5} 5
\times 4 + {c_{1} \over 4 } -\alpha c_{3}) + \right.
\nonumber
\\
\left. + \sum _{k=3,4,...}  c_{2k+1} (2k+1) (2k)  \xi^{2k-1}  + {1
\over 4}\sum _{k=1,2,...} c_{2k+1}\; \xi^{2k+3} -
 \alpha
\sum _{k=2,3,...} c_{2k+1}  \xi^{2k+1}  \right ] _{\mbox{odd}} =
0 \; .
\end{eqnarray}

\noindent
From this it follows
\begin{eqnarray}
\left [ \xi^{0} (2    c_{2} -\alpha c_{0} ) +\xi^{2} (c_{4} \; 4 \times 3  +
{c_{0} \over 4} - \alpha \; c_{2}) \; + \xi^{4}\; ( c_{6}\; 6\times 5 +
{c_{2} \over 4} - \alpha  c_{4} ) + \right.
\nonumber
\\
\left. + \sum _{n=3,4,...}  \left (  c_{2n+2} (2n+2) (2n+1) + {1
\over 4} \; c_{2n-2} - \alpha   \; c_{2n} \; \right ) \xi^{2n} \; \right
]_{\mbox{even}}\;+
\nonumber
\\
\left [ \;  \xi (c_{3} 3 \times 2  - \alpha  \; c_{1}) + \xi^{3} (c_{5} 5
\times 4 + {c_{1} \over 4 } - \alpha c_{3}) + \right.
 \nonumber
\\
\left. + \sum _{n=3,4,...}   \left ( c_{2n+1} (2n+1) (2n)    + {1
\over 4} \; c_{2n-3}  -
 \alpha  \; c_{2n-1} \right ) \;   \xi^{2n-1}  \right ] _{\mbox{odd}} = 0 \; .
\end{eqnarray}

\noindent Setting each coefficient at a $\xi^{k}$ equal to zero
one derives two independent groups of recurrent relations:
\underline{even}
\begin{eqnarray}
\left. \begin{array}{rl}
\xi^{0}:  &   \qquad  2 \;    c_{2} - \alpha  \; c_{0}  = 0 \; , \\[2mm]
\xi^{2}:  &   \qquad  c_{4} \; 4 \times 3  + {c_{0} \over 4} - \alpha \; c_{2}  =0 \; ,\\[2mm]
\xi^{4}:  &   \qquad  c_{6}\; 6\times 5 + {c_{2} \over 4} - \alpha  \; c_{4}  =0 \; , \\[2mm]
n=3,4,... , \;\; \xi^{2n}:  &  \qquad      c_{2n+2} (2n+2) (2n+1) +
{1 \over 4} \; c_{2n-2} - \alpha   \; c_{2n} \; = 0 \; ;
\end{array} \right.
\label{3.6b}
\end{eqnarray}
\underline{odd}
\begin{eqnarray}
\left. \begin{array}{rl}
\xi^{1}:  &  \qquad    c_{3} \; 3 \times 2  - \alpha \; c_{1} = 0 \; , \\[2mm]
\xi^{3}:  &  \qquad    c_{5} \; 5 \times 4 + {c_{1} \over 4 } - \alpha \; c_{3}=0 \; , \\[2mm]
n=3,4,... , \;\;  \xi^{2n-1}:  &   \qquad
 c_{2n+1} (2n+1) (2n)    + {1 \over 4} \; c_{2n-3}  -  \alpha \; c_{2n-1} = 0 \; .
 \end{array} \right.
\label{3.6c}
\end{eqnarray}

Taking into account  the absence of any connection of equations
(\ref{3.6b}) and (\ref{3.6c}) one can construct two linearly
independent solutions (even and odd respectively):
\underline{even}
\begin{eqnarray}
c_{0}=1, c_{1} =0 ,  \qquad  F_{1}(\xi)  = 1 + c_{2}\xi^{2} +
c_{4} \xi^{4} + ... , \label{3.7a}
\end{eqnarray}
\begin{eqnarray}
c_{2} = {\alpha  \over 2} \; , \qquad c_{4} = {1 \over 4 \times 3} \;
(\; \alpha \; c_{2} - {1 \over 4}\;)= {1 \over 4!  } \; (\; \alpha^{2} -{1 \over 2}\; )\; ,
\nonumber
\\
c_{6} = {1 \over 6 \times 5} \; (\; \alpha \; c_{4} - { c_{2} \over
4}\; ) = {1 \over 6!} \; (\; \alpha^{3} - {7\over 2}\; ) \; , \qquad
\nonumber
\\
n=3,4,... : \qquad c_{2n+2}  =  {1 \over (2n+2) (2n+1)} \; ( \; \alpha
\; c_{2n} \; - \; {1 \over 4} \; c_{2n-2} \; )
\end{eqnarray}
\underline{odd}
\begin{eqnarray}
c_{0}=0, c_{1} =1, \qquad  F_{2}(\xi) = \xi + c_{3}\xi^{3} +
c_{5} \xi^{5} + ... \; ,
\nonumber
\\
c_{3} = {1 \over 3!} \; \alpha  \; , \qquad c_{5} = { 1 \over 5 \times
4}\; (\; \alpha \; c_{3} -{c_{1}\over 4}\;)= {1 \over 5!}\; (\alpha ^{2}
-{3\over 2}\; ) \; ,
\nonumber
\\
n=3,4,... : \qquad    c_{2n+1} = {1 \over (2n+1) (2n) } \; ( \;  \alpha
\; c_{2n-1}  \;- \; {1 \over 4} \; c_{2n-3} \; ) \; . \label{3.7b}
\end{eqnarray}
or differently

\underline{even}
\begin{eqnarray}
F_{1}(\xi^{2}) = 1 + a_{2}{\xi^{2} \over 2!}  + a_{4} {\xi^{4}\over 4!}  +
... , \label{3.8a}
\\
a_{2} = \alpha  \; , \qquad a_{4} =  \alpha^{2} -{1 \over 2}\; , \qquad
c_{6} =  \alpha^{3} - {7\over 2} \alpha \; ,
\nonumber
\\
n=3,4,... : \qquad a_{2n+2}  =  \alpha  \; a_{2n}  -  {(2n)(2n-1) \over
4}\; a_{2n-2} \; \; ;
\nonumber
\end{eqnarray}

\underline{odd}
\begin{eqnarray}
F_{2} (\xi) = \xi + a_{3}\; {\xi^{3} \over 3!}  + a_{5}\;{ \xi^{5} \over
5!} + ... \; ,
\nonumber
\\
a_{3} =  \alpha  \; , \qquad a_{5} = \alpha^{2} - {3\over 2} \; ,
\nonumber
\\
n=3,4,... : \qquad  a_{2n+1} =   \alpha  \; a_{2n-1}  \;- \;
{(2n-1)(2n-2) \over 4}\; a_{2n-3} \;  \; .
\label{3.8b}
\end{eqnarray}

%4
\section{ The set of basis wave functions  for Klein-Fock  particle,
the role and manifestation of vector and spinor space structures respectively}

\hspace{5mm}
Having combined
two previous solutions $F_{1}$ and $F_{2}$, we can obtain four types
of the wave functions, solutions of the Klein-Fock equation in cylindrical
parabolic coordinates (we will change the notation: $F_{1} \Longrightarrow E; \; F_{1} \Longrightarrow  O$ :

%Таким образом, возможны четыре типа решений уравнения Клейна-Фока в цилиндрических
%параболических координатах (экспоненциальные множители
%$e^{-i\epsilon t / \hbar} $  и $\; e^{ipz/\hbar }$ опускаем)
\begin{eqnarray}
\left. \begin{array}{ll}
(\mbox{even} \otimes \mbox{even}): \qquad &
\Phi_{++} =   E(a, u^{2}) \; E(-a, v^{2})   ,\\[1mm]
(\mbox{odd} \otimes \mbox{odd)}:   \qquad &
\Phi_{--} = O (a, u) \;  O(-a, v\;) \;\; ,\\[1mm]
(\mbox{even} \otimes \mbox{odd)}:   \qquad &
\Phi_{+-} =  E (a, u^{2}) \;  O(-a, v\;) \; ,\\[1mm]
(\mbox{odd} \otimes \mbox{even}):   \qquad &
\Phi_{-+} =  O (a, u) \; E(-a, v^{2}) \; .
\end{array} \right.
\label{4.1}
\end{eqnarray}

Having  in mind relation   between $(u,v)$ and $(x,y)$,  one readily notes behavior
of the wave functions constructed at the point $x=0,y=0$ (variable $z$ is omitted):
\begin{eqnarray}
\left. \begin{array}{ll} (\mbox{even} \;\;\otimes \;\;
\mbox{even}): \qquad &
\Psi_{++}(x=0,y=0) \neq 0 \; , \\
(\mbox{odd} \otimes \mbox{odd}): \qquad &
\Psi_{--}(x=0,y=0) =  0 \; , \\
(\mbox{even} \otimes \mbox{odd}): \qquad &
\Psi_{+-}(x>0,y=0) =  0 \; , \\
(\mbox{odd} \otimes \mbox{even}): \qquad &
\Psi_{-+}(x<0,y=0) =  0 \;.
\end{array} \right.
\label{4.3}
\end{eqnarray}

Now let us consider which restrictions for the wave functions $\Psi$
are imposed by the requirement of single-valued ness.
Two peculiarities  in parameterizing are substantial:
\begin{eqnarray}
\underline{v=0\;:}  \qquad  x = + {u^{2} \over 2} \geq 0 , \; y =0 ; \qquad
\underline{u=0\; : } \qquad x= - {v^{2} \over 2}  \leq 0 , \; y =0  \; .
\label{4.4}
\end{eqnarray}

%Эти соотношения можно пояснить рисунками
\vspace{-3mm}

\unitlength=0.35mm
\begin{picture}(100,50)(-80,0)
\special{em:linewidth 0.4pt} \linethickness{0.6pt}

\put(-50,0){\vector(+1,0){100}}  \put(65,-5){$x$}
\put(0,-50){\vector(0,+1){100}}  \put(-10, +45){$y$}

\put(0,+1){\line(+1,0){50}}  \put(0,+0.5){\line(+1,0){50}}
\put(0,+1.3){\line(+1,0){50}}  \put(0,-0.3){\line(+1,0){50}}

\put(+100,0){\vector(+1,0){100}}  \put(215,-5){$x$}
\put(150,-50){\vector(0,+1){100}}  \put(140, +45){$y$}

\put(150,+1){\line(-1,0){50}}  \put(150,+0.5){\line(-1,0){50}}
\put(150,+1.3){\line(-1,0){50}}  \put(150,-0.3){\line(-1,0){50}}

\end{picture}

\vspace{20mm}

\begin{center}
{\bf Fig 4. The peculiarities in parametrization
%Особенности отображения $(x,y) \Longrightarrow (u,v)$
}
\end{center}

%\vspace{10mm}

The above four solutions (\ref{4.1}) behave in peculiar regions as follows:
\vspace{2mm}
%\vspace{25mm}
$
\underline{(\mbox{even} \otimes \mbox{even}):}
$
\begin{eqnarray}
 \Phi_{++}(a;u=0,v) =   E (a, u^{2}=0)  E (-a, v^{2}) =
\nonumber
\\
=E(-a, v^{2}) = + \; \Phi_{++}(a;u=0,-v) \; ,
\nonumber
\\
 \Phi_{++}(a;+u,v=0) =   E (a, u^{2})  \;E(-a, v^{2}=0) =
\nonumber
\\
= E (+a, u^{2})=  + \;\Phi_{++}(a;-u,v=0) \; ,
\nonumber
\end{eqnarray}

\underline{$(\mbox{odd} \otimes \mbox{odd)}:$}
\begin{eqnarray}
 \Phi_{--}(a;u =0,+v) = O (a, u=0) \; O(-a, v\;)  =
\nonumber
\\
=  + \; \Phi_{--}(a;u=0,-v) = 0
\nonumber
\\
\Phi_{--}(a;u,v=0) = O (+a, u) \; )O(-a, v=0\;)  =
\nonumber
\\
=  +  \;
\Phi_{--}(a;-u,v=0)  = 0 \; ,
\nonumber
\end{eqnarray}

\underline{$ (\mbox{even} \otimes \mbox{odd)}:$}
\begin{eqnarray}
 \Phi_{+-}(a;u=0,+v) =  E (a, u^{2}=0)  \; O(-a, v\;) =
\nonumber
\\
=O(-a, v\;) = - \;O(-a, -v\;)= -\; \Phi_{+-}(a;u=0,-v)\; ,
\nonumber
\\
 \Phi_{+-}(a;u,v=0) =  E (+a, u^{2})  \; O(-a, v=0\;) =
\nonumber
\\
=  \Phi_{+-}(a;-u,v=0) = 0 \; ,
\nonumber
\end{eqnarray}

\underline{$(\mbox{odd} \otimes \mbox{even}):$}
\begin{eqnarray}
\Phi_{-+}(a;u=0,+ v) =  O (+a, u=0) \;  E(-a, v^{2}) =
\nonumber
\\
= \Phi_{-+}(a;u=0,- v)  = 0 \; ,
\nonumber
\\
\Phi_{-+}(a;+u,v=0) =  O (+a, u) \;  E(-a, v^{2}=0) =
\nonumber
\\
= O (+a, u) = - O (a, -u) = - \; \Phi_{-+}(a;-u,v=0) \; .
\label{4.5}
\end{eqnarray}

 Taking in mind the Fig.1 and the Fig. 3, one cam immediately conclude:
solutions $\Phi$ of the types   $(++)$ and $(--)$ are
single-valued in the space with vector structure, whereas the
solutions of the types  $(+-)$ and $(-+)$
are not single-valued in space with vector structure,
so  these types  $(+-)$ and $(-+)$  must be rejected.
However, these solutions $(+-)$ and $(-+)$ must be retained in the space with spinor structure.

That dividing of the basis wave functions into two subsets
may be formalized mathematically with the help of special discrete operator acting
in spinor space:
\begin{eqnarray}
\hat{\delta} = \left | \begin{array}{cc}
-1 & 0 \\ 0  & -1  \end{array} \right | , \qquad \hat{\delta}
\left | \begin{array}{c} u \\ v
\end{array} \right | =
\left | \begin{array}{c} -u \\ -v
\end{array} \right | \; .
\label{4.6a}
\end{eqnarray}

It is easily verified that  solutions single-valued in  the vector space model
 are eigenfunctions of $\delta$ with eigenvalue $\delta = +1$:
\begin{eqnarray}
 \hat{\delta} \; \Phi_{++}(a;u,v) = + \; \Phi_{++}(a;u,v) \; ,
\;\;
\hat{\delta} \; \Phi_{--}(a;u,v) = + \; \Phi_{--}(a;u,v) \; ,
\label{4.6b}
\end{eqnarray}

\noindent and additional ones acceptable
only in  the spinor space model, are eigenfunction with the eigenvalue  $\delta = -1$:
\begin{eqnarray}
 \hat{\delta} \; \Phi_{+-}(a;u,v) = - \; \Phi_{+-}(a;u,v) \; ,
\;\;
\hat{\delta} \; \Phi_{-+}(a;u,v) = - \; \Phi_{-+}(a;u,v) \; .
\label{4.6c}
 \end{eqnarray}

When using the spinor space model,  two set
$(u,v)$ and $(-u,-v)$  represent different geometrical points in the spinor space, so  the
requirement of single valuedness as applied in the case of spinor space does
not presuppose that the values of the wale functions must be equal in the points
$(u,v)$ and $(-u,-v)$:
\begin{eqnarray}
 \Phi (u,v) = \Phi((x,y)^{(1)}) \neq
 \Phi (-u,-v) = \Phi((x,y)^{(2)})
 \;  .
\label{4.6d}
\end{eqnarray}

\vspace{5mm}

Now let us add some details more. In general, the vector plane  $(x,y)$
allows three inversion operations to which one can relate  six discrete operations
in spinor "plane" \hspace{2mm}  $(u,v)$:
\begin{eqnarray}
(x,y)   \Longrightarrow   (x,-y) , \qquad
\hat{\pi} = \left | \begin{array}{cc}
+1 & 0 \\ 0  & -1 \end{array} \right | , \qquad
\hat{\pi'} =  \hat{\delta} \; \hat{\pi}  =  - \hat{\pi} \; ,
\nonumber
\\
 (x,y) \Longrightarrow (-x,y) , \qquad \hat{\omega} = \left | \begin{array}{cc}
0 & +1 \\ +1  & 0 \end{array} \right | , \qquad
\hat{\omega '} =  \hat{\delta} \; \hat{\omega}  =  - \hat{\omega} \;,
\nonumber
\\
(x,y) \Longrightarrow (-x,-y),
 \qquad \hat{R} =   \left | \begin{array}{cc}
0 & -1 \\ +1  & 0 \end{array} \right | , \qquad
\hat{R '} =  \hat{\delta} \; \hat{R}  =  - \hat{R}\; .
\label{4.7}
\end{eqnarray}

One can easily construct  eigenfunctions of these discrete
operations (\ref{4.7}) as well. For instance, let us consider the
operator $\hat{R} = \hat{\omega} \;\hat{\pi} $. Noting two
identities
\begin{eqnarray}
\hat{R} \;\Phi_{++}(a;u,v) = \hat{R} \;\;  E (a,u^{2}) \; E(-a,v^{2}) =
\nonumber
\\
= E (a,v^{2}) \; E(-a,u^{2})= \Phi_{++}(-a;u,v) \; ,
\label{4.8a}
\\
\hat{R} \; \Phi_{--}(a;u,v) = \hat{R} \;\; O(a,u)  \; O(-a,v) =
\nonumber
\\
=
O (a,-v) \; O(-a,u)= - \Phi_{--}(-a;u,v)\; .
\label{4.8b}
\end{eqnarray}

\noindent
one can easily construct the eigen-functions  of the operator $\hat{R}$
(arguments are omitted):
\begin{eqnarray}
\Phi_{++}^{(R= \pm 1)}=
\Phi_{++}(a) \; \pm \; \Phi_{++}(-a) \; ,\;\;
\hat{R} \; \; \Phi_{++}^{(R= \pm 1)} = \pm \;
\Phi_{++}^{(R=\pm 1)} \; ;
\label{4.9a}
\end{eqnarray}
\noindent
and
\begin{eqnarray}
\Phi_{--}^{(R = \pm 1)} =
\Phi_{--}(a) \; \mp \; \Phi_{--}(-a) \; ,
\;\;
\hat{R} \; \Phi_{--}^{(R=\pm 1)}  = \pm  \;
\Phi_{--}^{(R=\pm 1)}  \; .
\label{4.9b}
\end{eqnarray}

In the same way, taking into account the identities
\begin{eqnarray}
\hat{R} \; \Phi_{+-}(a;u,v) = \hat{R} \; E (a,u^{2})\;  O(-a,v) =
\nonumber
\\
=E(a,v^{2}) \;  O(-a,u)= + \; \Phi_{-+}(-a;u,v)
\label{4.10a}
\end{eqnarray}

\noindent
and
\begin{eqnarray}
\hat{R} \;  \Phi_{-+}(a;u,v) = \hat{R} \;  O (a,u) \; E(-a,v^{2}) =
\nonumber
\\
=O (a,-v) \; E(-a,u^{2})= - \; \Phi_{+-}(-a;u,v) \; ,
\label{4.10b}
\end{eqnarray}

\noindent
one can easily construct eigenfunctions with
complex eigenvalues:
\begin{eqnarray}
\varphi^{(R= \mp i)}   = \Phi_{+-}(a) \pm i \;
\Phi_{-+}(-a) \; , \;\;
\hat{R} \; \varphi^{(R= \pm i)}   = \pm \; i  \; \varphi^{(R= \pm i)}   \; ;
\label{4.11a}
\end{eqnarray}

\noindent
and
\begin{eqnarray}
\varphi^{(R=\mp i)} (-a)  = \Phi_{+-}(-a) \pm i \;
\Phi_{-+}(+a) \; ,\;\;
\hat{R} \; \varphi^{(\pm i)} (-a)  = \pm \; i  \; \varphi^{(\mp i)} (-a)  \; .
\label{4.11b}
\end{eqnarray}

Thus, there exist quite a definite  classification of the Klein-Fock  solutions
in cylindrical parabolic coordinates in terms of quantum numbers,
eigenvalues of the following operator
(an explicit form $\hat{A}$ will be given below)
\begin{eqnarray}
i {\partial \over  \partial t} \Longrightarrow \epsilon \; , \;\;
-i {\partial \over  \partial z} \Longrightarrow p  \; , \qquad
\hat{A} \Longrightarrow a \; , \;     (\hat{\delta} , \hat{R})
\Longrightarrow ( \delta  =  \pm 1, R = \pm 1 ) \; . \label{4.12}
\end{eqnarray}

As a base to classify solutions of the Klein-Fock equation, instead of $(\hat{\delta} , \hat{R})$
 one might have taken other
two operator: for instance, $\hat{\delta}$ and $\hat{\omega}$.
Then, allowing for the  identities
\begin{eqnarray}
\hat{\omega} \;\Phi_{++}(a;u,v) = \hat{\omega} \;\;  E (a,u^{2}) \; E(-a,v^{2}) =
\nonumber
\\
= E (a,v^{2}) \; E(-a,u^{2})= \Phi_{++}(-a;u,v) \; ,
\label{4.13a}
\\
\hat{\omega} \; \Phi_{--}(a;u,v) = \hat{\omega} \;\; O(a,u)  \; O(-a,v) =
\nonumber
\\
=
O (a,v) \; O(-a,u)=  \Phi_{--}(-a;u,v)\; .
\label{4.13b}
\end{eqnarray}

We can construct eigenfunctions of the operator $\hat{\omega}$:
\begin{eqnarray}
\Phi_{++}^{(\omega = \pm 1)}=
\Phi_{++}(a) \; \pm \; \Phi_{++}(-a) \; ,
\;\;
\hat{\omega } \; \; \Phi_{++}^{(\omega = \pm 1)} = \pm \;
\Phi_{++}^{(\omega =\pm 1)} \; ;
\label{4.14a}
\end{eqnarray}

\noindent
and
\begin{eqnarray}
\Phi_{--}^{(\omega  = \pm 1)}=
\Phi_{--}(a) \; \pm \; \Phi_{--}(-a) \; ,
\;\; \hat{\omega} \; \Phi_{--}^{(\omega =\pm 1)} = \pm  \;
\Phi_{--}^{(\omega =\pm 1)}) \; .
\label{4.14b}
\end{eqnarray}

In the same manner, for additional solutions we have
\begin{eqnarray}
\hat{\omega} \; \Phi_{+-}(a;u,v) = \hat{\omega} \; E (a,u^{2})\;  O(-a,v) =
\nonumber
\\
= E (a,v^{2}) \;  O(-a,u)= + \; \Phi_{-+}(-a;u,v)
\label{4.15a}
\end{eqnarray}

\noindent
and
\begin{eqnarray}
\hat{\omega} \;  \Phi_{-+}(a;u,v) = \hat{\omega } \;  O (a,u) \; E(-a,v^{2}) =
\nonumber
\\
=O (a,v) \; E(-a,u^{2})=  \; \Phi_{+-}(-a;u,v) \; ,
\label{4.15b}
\end{eqnarray}

\noindent
therefore, the eigenfunctions  may be given as
\begin{eqnarray}
\varphi^{(\omega = \pm 1)}   = \Phi_{+-}(a) \pm  \; \Phi_{-+}(-a) \; ,
\;\;
\hat{\omega } \; \varphi^{(\omega = \pm 1)}   =
 \pm \;  \varphi^{(\omega = \pm 1)}   \; ;
\label{4.16a}
\\
\varphi^{(\omega = \pm 1 )}   = \Phi_{+-}(-a) \pm \;
\Phi_{-+}(+a) \; ,
\;\;
\hat{\omega} \; \varphi^{(\omega = \pm 1)}   = \pm \;  \varphi^{(\omega = \mp 1)}   \; ;
\label{4.16b}
\end{eqnarray}

It is easy to obtain some classifications with the help of
$(\hat{\delta},\; \hat{\pi})$. Indeed,
\begin{eqnarray}
\hat{\pi} \; \Psi _{++}(a;u,v) = \hat{\pi}\; F_{1}(a,u^{2}) \; F_{1}(-a,v^{2}) =
F_{1}(a,u^{2}) \; F_{1}(-a,v^{2}) = + \; \Psi _{++}(a;u,v)  \; ,
\nonumber
\\
\hat{\pi} \; \Psi _{--}(a;u,v) = \hat{\pi}\; F_{2}(a,u) \; F_{2}(-a,v) =
F_{2}(a,-u) \; F_{2}(-a,v) = - \; \Psi _{--}(a;u,v)  \; ,
\nonumber
\\
\hat{\pi} \; \Psi _{+-}(a;u,v) = \hat{\pi}\; F_{1}(a,u^{2}) \; F_{2}(-a,v) =
F_{1}(a,u^{2}) \; F_{2}(-a,v) = + \; \Psi _{+-}(a;u,v)  \; ,
\nonumber
\\
\hat{\pi} \; \Psi _{-+}(a;u,v) = \hat{\pi}\; F_{2}(a,u) \; F_{1}(-a,v^{2}) =
F_{2}(a,-u) \; F_{1}(-a,v^{2}) = - \; \Psi _{-+}(a;u,v)  \; .
%\nonumber
%\\
\label{4.17}
\end{eqnarray}

\begin{quotation}
\noindent
Remembering eqs. (\ref{4.6a}) --  (\ref{4.6d}), one can conclude that the basic solutions
are eigenfunctions of two discrete operators
 $\hat{\delta}$ and $\hat{\pi}$:
\begin{eqnarray}
(\Psi_{++},\; \Psi_{--},\; \Psi_{+-},\; \Psi_{-+} ) \qquad \Longleftrightarrow  \qquad (\hat{\delta}, \; \hat{\pi})
\nonumber
\end{eqnarray}

\noindent All three ways to classify solutions with the help of discrete operators
$(\hat{\delta}, \hat{\pi}) , \; (\hat{\delta}, \hat{R}), \;
(\hat{\delta}, \hat{\omega}) \;
$
are equally acceptable. It is understandable that an operator
$\hat{A}$
related to the quantum number $a$, must commute with
$\hat{\delta}$ and $ \hat{\pi}$, and
it is not commute with  $\hat{R}$ and $ \hat{\omega}$.

Boundary properties of the wave functions constructed can be illustrated by the schemes:

%\vspace{-47mm}

\unitlength=0.33mm
\begin{picture}(100,100)(-100,0)
\special{em:linewidth 0.4pt} \linethickness{0.6pt}

\put(-70,+50){$\underline{\Psi_{+\;+}}$}
\put(-50,0){\vector(+1,0){100}}  \put(65,-5){$x$}
\put(0,-50){\vector(0,+1){100}}  \put(-10, +45){$y$}
\put(+20,+20){\vector(-1,-1){18}} \put(+24,+20){non-zero} \put(0,0){\circle*{4}}

\put(+100,+50){$\underline{\Psi_{-\;-}}$}
\put(+150,0){\vector(+1,0){100}}  \put(265,-5){$x$}
\put(200,-50){\vector(0,+1){100}}  \put(+190, +45){$y$}
\put(+220,+20){\vector(-1,-1){18}} \put(+224,+20){zero} \put(200,0){\circle*{4}}

\put(-70,-80){$\underline{\Psi_{+\;-}}$}
\put(-50,-130){\vector(+1,0){100}}  \put(65,-140){$x$}
\put(0,-180){\vector(0,+1){100}}  \put(-10, -90){$y$}
\put(+40,-110){\vector(-1,-1){18}} \put(+44,-127){zero} \put(0,-131){\circle*{4}}
\put(0,-131){\line(+1,0){50}}
%\put(0,-151.5){\line(+1,0){50}}
\put(-50,-128){$+\; + \; + \; + $}
\put(-50,-138){$-\; - \; - \; - $}

\put(+100,-80){$\underline{\Psi_{-\;+}}$}
\put(+150,-130){\vector(+1,0){100}}  \put(265,-135){$x$}
\put(200,-180){\vector(0,+1){100}}  \put(+190, -90){$y$}
\put(+160,-110){\vector(+1,-1){18}} \put(+140,-120){zero} \put(200,-131){\circle*{4}}
\put(200,-131){\line(-1,0){50}}  \put(200,-131.5){\line(-1,0){50}}
\put(+200,-128){$+\; + \; + \; + $}
\put(+200,-138){$-\; - \; - \; - $}

\end{picture}

\vspace{57mm}

\begin{center}
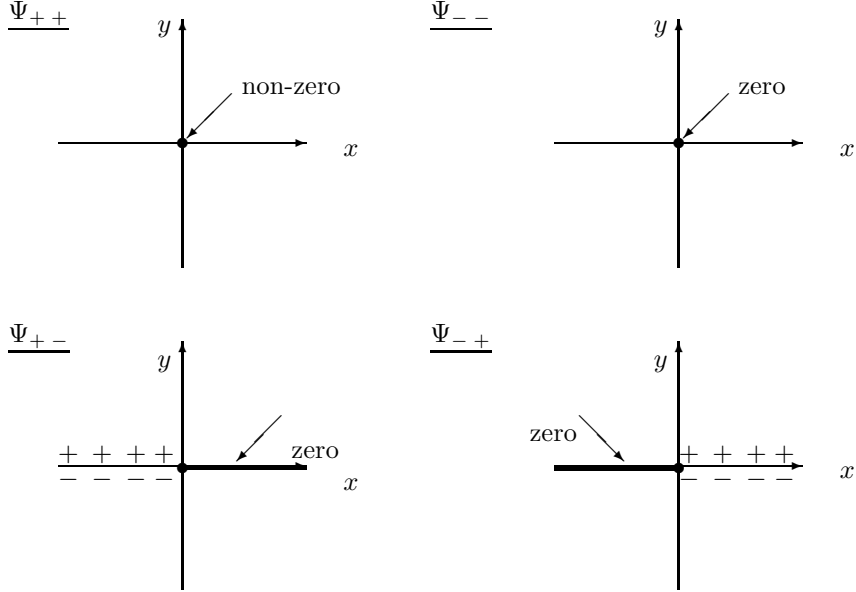
{\bf Fig 7. Boundary behavior of  the wave functions in  $(x,y)$-plane}
\end{center}

\end{quotation}

%5
\section{Explicit form of a diagonalized operator $\hat{A} $ }

\hspace{5mm}
Let us find an explicit, form of the operator $\hat{A}$
introduced above by the equation $\hat{A} \Psi = a \;\Psi $.
To this end, remembering eq. (\ref{3.2})
\begin{eqnarray}
{d^{2} U \over  d u^{2}} +   \lambda^{2}  \; u^{2} \; U =   a \; U \; , \qquad
{d^{2} V \over  d v^{2}} +  \lambda^{2} \; v^{2}\;   V =  - a \;   V \; ,
\label{5.1a}
\end{eqnarray}

\noindent one derives
\begin{eqnarray}
{1 \over 2 }  \;
\left [ ({\partial^{2}  \over  \partial u^{2}} +   \lambda^{2}  \; u^{2} ) -
({\partial^{2}  \over  \partial  v^{2}} +  \lambda^{2} \; v^{2} ) \right  ] \; U(u) V(v) = a \; U(u)
\;V (v) \; .
\label{5.1b}
\end{eqnarray}

\noindent
From where one gets an  explicit form of $\hat{A}$:
\begin{eqnarray}
\hat{A} \; \Psi _{\epsilon, p, a} (t,u,v,z)  = a \; \Psi _{\epsilon, p, a} (t,u,v,z)  \; ,
\label{5.1c}
\end{eqnarray}

\noindent and
\begin{eqnarray}
\hat{A} =
{1 \over 2 }  \;
\left \{  \left [ \; {\partial ^{2}  \over  \partial u^{2}} +   ( \;
-  {\partial^{2} \over \partial  t^{2}}
+ {\partial^{2} \over  \partial z^{2}} -  m^{2}  )
\; u^{2} \; )\;  \right ]  -
 \left [ \;{\partial^{2}  \over  \partial v^{2}} +
( \; -  {\partial^{2} \over \partial  t^{2}}
+ {\partial^{2} \over  \partial z^{2}} - m^{2}  \; )
 \; v^{2} \; \right ]  \right  \} \; .
\nonumber
\end{eqnarray}

Let us transform this operator  $\hat{A}$
to Cartesian coordinates.  To this end, taking into account the  formulas
\begin{eqnarray}
{\partial \over \partial u} = u \; {\partial \over \partial x }  + v \;{\partial \over \partial y} ,
\qquad
{\partial \over \partial v} = -v  \;{\partial \over \partial x }  + u \;{\partial \over \partial y} ,
\nonumber
\\
{\partial ^{2} \over  \partial u^{2}} =
u {\partial \over \partial x} u {\partial \over \partial x} +
u {\partial \over \partial x} v {\partial \over \partial y} +
v {\partial \over \partial y} u {\partial \over \partial x} +
v {\partial \over \partial y} u {\partial \over \partial y} \; ,
\nonumber
\\
{\partial ^{2} \over  \partial v^{2}} =
v {\partial \over \partial x} v {\partial \over \partial x} -
v {\partial \over \partial x} u {\partial \over \partial y} -
u {\partial \over \partial y} v {\partial \over \partial x} +
u {\partial \over \partial y} u {\partial \over \partial y} \; ;
\nonumber
\end{eqnarray}

\noindent
and also
\begin{eqnarray}
{\partial u \over  \partial x} = { u \over u^{2} + v^{2}} \; , \qquad
{\partial u \over  \partial y} = { v \over u^{2} + v^{2}} \; ,
\qquad
{\partial v \over  \partial x} = { -v \over u^{2} + v^{2}} \; , \qquad
{\partial v \over  \partial y} = { u \over u^{2} + v^{2}} \; ,
\nonumber
\end{eqnarray}

\noindent one finds
\begin{eqnarray}
{\partial ^{2} \over  \partial u^{2}} =
u^{2} {\partial^{2} \over  \partial x^{2}} + v^{2} {\partial^{2} \over \partial y^{2}} +
2uv \; {\partial ^{2} \over \partial x \partial y} + {\partial \over \partial x} \; ,
\nonumber
\\
{\partial ^{2} \over  \partial v^{2}} =
v^{2} {\partial^{2} \over  \partial x^{2}} + u^{2} {\partial^{2} \over \partial y^{2}} -
2uv \; {\partial ^{2} \over \partial x \partial y} - {\partial \over \partial x} \; ,
\nonumber
\end{eqnarray}

\noindent
that is
\begin{eqnarray}
{1 \over 2} \; ( {\partial ^{2} \over  \partial u^{2}}-
{\partial ^{2} \over  \partial v^{2}} )
   =
{u^{2} - v^{2} \over 2} \;   ({\partial ^{2} \over \partial x^{2}}
- {\partial ^{2} \over \partial y^{2}} )
 + 2uv \; {\partial ^{2} \over \partial x \partial y}  +  {\partial \over \partial x} =
\nonumber
\\
= x  \;   ({\partial ^{2} \over \partial x^{2}}
- {\partial ^{2} \over \partial y^{2}} )
 + 2y \; {\partial ^{2} \over \partial x \partial y}  +  {\partial \over \partial x} \; .
\label{5.2}
\end{eqnarray}

Therefore, for $\hat{A}$ in Cartesian coordinates one  has the following
representation
\begin{eqnarray}
\hat{A} =
x  \;   ({\partial ^{2} \over \partial x^{2}}
- {\partial ^{2} \over \partial y^{2}} )
 + 2y \; {\partial ^{2} \over \partial x \partial y}  +  {\partial \over \partial x}
+ x \;
( \;
-  {\partial^{2} \over \partial  t^{2}}
+ {\partial^{2} \over  \partial z^{2}} -  m^{2} ) \; ;
\label{5.3a}
\end{eqnarray}

\noindent whereas in $(u,\vartheta,z)$-coordinates it looks as
\begin{eqnarray}
\hat{A} = {1 \over 2} \; \left [ \;
({\partial^{2} \over  \partial u^{2}} - {\partial^{2} \over  \partial v^{2}} ) -
( {\partial^{2} \over  \partial t^{2}} - {\partial^{2} \over  \partial z^{2}} + m^{2} )
\; (u^{2} - v^{2}) \; \right ] \; .
\label{5.3b}
\end{eqnarray}

The solutions constructed above,
$\Psi_{++},\Psi_{--},\Psi_{+-},\Psi_{-+}$, behave themselves in exact correspondence
with the following commutation relations:
\begin{eqnarray}
\hat{\delta} \; \hat{A} =  + \; \hat{A}  \;\hat{\delta} \; , \qquad
\hat{\pi} \; \hat{A} =  + \; \hat{A}  \;\hat{\pi} \; ,
\nonumber
\\
\hat{\omega} \; \hat{A} =  - \; \hat{A}  \;\hat{\omega} \; , \qquad
\hat{R} \; \hat{A} =  - \; \hat{A}  \;\hat{R} \; ,
\nonumber
\\
\hat{\pi} \; \hat{\omega} = - \;  \hat{\omega}\;  \hat{\pi} \; , \qquad
\hat{\pi} \; \hat{R} = -  \; \hat{R}\;  \hat{\pi} \; .
\label{5.4}
\end{eqnarray}

%6
\section{Orthogonality  and completeness
of the bases for  vector and spinor space models }

\hspace{5mm}
Now let us consider the scalar multiplication
\begin{eqnarray}
\int \Psi^{*}_{\mu'} \; \Psi _{\mu} \; \sqrt{-g}  \; dt dz  du dv \; .
\label{6.1a}
\end{eqnarray}

\noindent
of the basic wave functions constructed:
\begin{eqnarray}
\Psi_{++}(\epsilon, p, a) = e^{i\epsilon t} \; e^{ipz} \;
\Phi_{++}(a;u,v) =  e^{i\epsilon t} \; e^{ipz} \;  E (+a, u^{2}) \; E(-a, v^{2})  \; ,
\nonumber
\\
\Psi_{--}(\epsilon, p, a) = e^{i\epsilon t} \; e^{ipz} \;
\Phi_{--}(a;u,v) =e^{i\epsilon t} \; e^{ipz} \;  O (+a, u) \;  O(-a, v\;) \; ,
\nonumber
\\
\Psi_{+-}(\epsilon, p, a) = e^{i\epsilon t} \; e^{ipz} \;
\Phi_{+-}(a;u,v) =  e^{i\epsilon t} \; e^{ipz} \; E (+a, u^{2}) \;  O(-a, v\;) \; ,
\nonumber
\\
\Psi_{-+}(\epsilon, p, a) = e^{i\epsilon t} \; e^{ipz} \;
\Phi_{-+}(a;u,v) = e^{i\epsilon t} \; e^{ipz} \;  O (+a, u) \; E(-a, v^{2}) \; .
\label{6.1b}
\end{eqnarray}

\vspace{5mm}
\noindent $\mu$ and $\mu'$ stand  for generalized quantum numbers.
In the first place, interesting integrals are (arguments (a;u,v) are omitted):

\underline{in vector space}
\begin{eqnarray}
I_{0} =
\int_{0} ^{+\infty} dv   \int _{-\infty}^{+\infty} du
\; \Phi_{++}^{*} \; \Phi_{--}\;  ( u^{2} + v^{2})  ,
\label{6.2a}
\end{eqnarray}

\underline{in spinor space}
\begin{eqnarray}
I_{1} =
\int_{-\infty} ^{+\infty} dv  \int _{-\infty} ^{+\infty} du
\; \Phi_{++}^{*}\; \Phi_{--}\;  ( u^{2} + v^{2})  \; ,
\nonumber
\\
I_{2} =
\int_{-\infty} ^{+\infty} dv  \int _{-\infty} ^{+\infty} du
\; \Phi_{+-}^{*} \; \Phi_{-+} \;  ( u^{2} + v^{2})  \; ,
\nonumber
\\
I_{3} =
\int_{-\infty} ^{+\infty} dv  \int_{-\infty} ^{+\infty} du
 \; \Phi_{++}^{*} \; \Phi_{+-} \;  ( u^{2} + v^{2}) \; ,
\nonumber
\\
I_{4} =
\int_{-\infty} ^{+\infty} dv   \int_{-\infty} ^{+\infty} du
 \; \Phi_{++}^{*} \; \Phi_{-+} \;  ( u^{2} + v^{2})  \; ,
\nonumber
\\
I_{5} =
\int _{-\infty} ^{+\infty} dv  \int_{-\infty} ^{+\infty} du
 \; \Phi_{--}^{*} \; \Phi_{+-} \;  ( u^{2} + v^{2}) \; ,
\nonumber
\\
I_{6}
 =
\int_{-\infty} ^{+\infty}  dv  \int _{-\infty} ^{+\infty} du
 \; \Phi_{--}^{*} \; \Phi_{-+}) \;  ( u^{2} + v^{2})  \; .
\label{6.2b}
\end{eqnarray}

\noindent
Integral $I_{0}$ in vector  space vanishes identically
\begin{eqnarray}
I =
\int_{0} ^{+\infty} dv   \int _{-\infty}^{+\infty} du
\; E (+a, u^{2}) \; E(-a, v^{2}) \times
O (+a, u) \;  O(-a, v\;) \;  ( u^{2} + v^{2}) =
\nonumber
\\
=
\int_{0} ^{+\infty} dv   \int _{-\infty}^{+\infty} du \;
E (+a, u^{2}) \;  \underline{O (+a, u) } \times
E(-a, v^{2}) ; O(-a, v\;) \;  ( u^{2} + v^{2} ) \equiv 0 \; ,
\nonumber
\end{eqnarray}

\noindent
because integration in variable  $u \in (- \infty, + \infty) $
is done for an odd function of $u$ in symmetrical region $u \in (- \infty, + \infty) $.
By the same reasons, integral  $I_{1}$ in spinor space  vanishes  as  well.

The integral $I_{2}$ vanishes
\begin{eqnarray}
I_{2} =
\int_{-\infty} ^{+\infty} dv  \int _{-\infty} ^{+\infty} du
\; \; E (+a, u^{2}) \;  O(-a, v\;)
 \;  O (+a, u) \; E(-a, v^{2})\;
   ( u^{2} + v^{2})  \equiv 0 \; ,
\nonumber
\end{eqnarray}

\noindent
because integration is done for an odd function in
$v,u$-variables, in symmetrical regions  $v \in (- \infty, + \infty)$ and
$u \in (- \infty, + \infty)$.

Integral $I_{3}$ vanishes
\begin{eqnarray}
I_{3} =
\int_{-\infty} ^{+\infty} dv  \int_{-\infty} ^{+\infty} du
 \;E (+a, u^{2}) \; E(-a, v^{2})
 \; E (+a, u^{2}) \;  O(-a, v\;) \;  u^{2} + v^{2}) \equiv 0 \; ,
\nonumber
\end{eqnarray}

\noindent
because integration is done for odd function of  $v$ variable, in the symmetrical  region
$v \in (- \infty, + \infty)$.

Integral  $I_{4}$ vanishes
\begin{eqnarray}
I_{4} =
\int_{-\infty} ^{+\infty} dv   \int_{-\infty} ^{+\infty} du
 \; E (+a, u^{2}) \; E(-a, v^{2})  \;
  O (+a, u) \; E(-a, v^{2})
  \;  ( u^{2} + v^{2})  \equiv 0 \; ,
\nonumber
\end{eqnarray}

\noindent
because integration is done for an odd function of U in symmetrical
region $u \in (- \infty, + \infty)$.

Integral $I_{5}$ vanishes
\begin{eqnarray}
I_{5} =
\int _{-\infty} ^{+\infty} dv  \int_{-\infty} ^{+\infty} du
 \; O (+a, u) \;  O(-a, v\;)
  \; E (+a, u^{2}) \;  O(-a, v\;) \;
 ( u^{2} + v^{2}) \equiv 0  \; ,
\nonumber
\end{eqnarray}

\noindent
because one integrates an odd  function  of  $u$ in symmetrical
region $u \in (- \infty, + \infty)$.

Integral $I_{6}$ vanishes
\begin{eqnarray}
I_{6}
 =
\int_{-\infty} ^{+\infty}  dv  \int _{-\infty} ^{+\infty} du
 \; O (+a, u) \;  O(-a, v\;)
  \; O (+a, u) \; E(-a, v^{2}) \;  ( u^{2} + v^{2}) \equiv 0  \; ,
\nonumber
\end{eqnarray}

\noindent
because one integrates an odd function  of  $v$ in symmetrical
region
$v \in (- \infty, + \infty)$.

\begin{quotation}

Thus, vanishing integrates $I_{0}, I_{1}... I_{6}$ from (\ref{6.2a}),(\ref{6.2b}) shows  that
the formulas  (\ref{6.1b}) provide us with orthogonal basis for Hilbert space $\Psi (unv,z)$,
where $(u,v,z)$
belong to an extended (spinor) space model.

\end{quotation}

%7
\section{On matrix elements of physical observables, in vector and  spinor space models}

\hspace{5mm}
The question of principle is how transition from vector to spinor space
model can
influence result  of calculation of matrix elements for physical quantities.
As an example, let us consider matrix elements for operator of coordinates:
One may calculate
matrix elements of basic initial coordinates $u, v$ or there 2-order derivative
coordinates $x,y$:
\begin{eqnarray}
(u,v) \qquad \mbox{or} \qquad  x = {u^{2} - v^{2} \over 2}\; ,
\; \; y = uv \;  .
\label{7.1}
\end{eqnarray}

With the use of the above  rules --  integral for an odd function in symmetrical
region  vanishes identically --  one can
derive simple section rules for matrix elements (for simplicity we restrict ourselves
only to the degeneracy in discrete quantum number $++,--,+-,-+$ taking $ \epsilon, p,  a$  fixed):

\underline{in vector space}
\begin{eqnarray}
\left. \begin{array}{ccc}
\underline{x_{\mu',\mu}}  \qquad  &     ++  &  -- \\[3mm]
++  &    \neq 0  &    0    \\
--  &   0 &    \neq 0
\end{array} \right.
, \qquad \qquad \qquad
\left. \begin{array}{ccc}
\underline{y_{\mu',\mu}}   \qquad &    ++  &   -- \\[2mm]
++  &  0 &    \neq 0     \\
--  &  \neq  0&    0
\end{array} \right.
\nonumber
\end{eqnarray}

\underline{in spinor space}
\begin{eqnarray}
\left. \begin{array}{ccccc}
\underline{x_{\mu',\mu} } \qquad &    ++  &    --  &    +-  &    -+   \\[3mm]
++            &     \neq 0    &     0  &     0  &     0   \\
--            &    0   &     \neq 0  &      0  &     0   \\
+-            &    0   &     0  &  \neq 0  &     0   \\
-+            &     0   &    0  &    0  &   \neq 0
\end{array} \right.
, \qquad\qquad
\left. \begin{array}{ccccc}
\underline{y_{\mu',\mu}} \qquad &    ++   &    --     &    +-  &   -+   \\[3mm]
++            &   0    &   \neq 0 &    0  &     0   \\
--          &    \neq 0    &    0 &     0  &    0   \\
+-            &     0    &    0     &    0  &    \neq 0   \\
-+            &   0    &    0     &   \neq 0  &     0
\end{array} \right.
\nonumber
\end{eqnarray}

The same for coordinates $u$ and $v$ looks:
\underline{in vector space}
\begin{eqnarray}
\left. \begin{array}{ccc}
\underline{u_{\mu',\mu}} \qquad  &    ++  & -- \\[2mm]
++  &    0 &   \neq 0    \\
--  &     \neq 0 &    0
\end{array} \right. ,
, \qquad \qquad \qquad
\left. \begin{array}{ccc}
\underline{v_{\mu',\mu} }  \qquad &    ++  &   -- \\[2mm]
++  &   \neq 0 &    0    \\
--  &   0 &    \neq 0
\end{array} \right.
\nonumber
\end{eqnarray}

\underline{in spinor space}
\begin{eqnarray}
\left. \begin{array}{ccccc}
\underline{u_{\mu',\mu} } \qquad  &    ++  &  --  &    +-  &  -+   \\[2mm]
++            &   0   &    0  &   0  &     \neq 0   \\
--            &   0   &    0  &  \neq 0  &    0   \\
+-            &   0   &  \neq 0  &    0  &    0   \\
-+       &  \neq 0   &     0  &   0  &    0
\end{array} \right.
, \qquad \qquad
\left. \begin{array}{ccccc}
\underline{v_{\mu',\mu}} \qquad   &   ++  &    --  &    +-  &    -+   \\[2mm]
++            &    0   &    0  &   \neq 0  &     0   \\
--            &     0   &   0  &    0  &   \neq 0   \\
+-            &    \neq 0   &    0  &    0  &    0   \\
-+            &     0   &    \neq 0  &     0  &    0
\end{array} \right.
\nonumber
\end{eqnarray}

Let us give some detail of calculation needed. Foe example,

\vspace{5mm} \underline{In vector space} $\qquad \qquad x_{a'++\;
,\;a--}$ =
\begin{eqnarray}
 =
\int_{0} ^{+\infty} dv   \int _{-\infty}^{+\infty} du
\;  E (+a', u^{2}) \; E(-a', v^{2})  \;
{u^{2} - v^{2} \over 2} \; O (+a, u) \;  O(-a, v\;)\; ( u^{2} + v^{2})  \equiv 0 \; ,
\nonumber
\end{eqnarray}

\underline{in spinor space} $\qquad \qquad  x_{a'++\; ,\;a--} =$
\begin{eqnarray}
 =
\int_{-\infty} ^{+\infty} dv   \int _{-\infty}^{+\infty} du
\;  E (+a', u^{2}) \; E(-a', v^{2})  \;
{u^{2} - v^{2} \over 2} \; O (+a, u) \;  O(-a, v\;)\; ( u^{2} + v^{2})  \equiv 0 \; .
\nonumber
\end{eqnarray}

\underline{In vector space} $\qquad \qquad u_{a'++\; ,\;a--}  =$
\begin{eqnarray}
 =
\int_{0} ^{+\infty} dv   \int _{-\infty}^{+\infty} du
\;  E (+a', u^{2}) \; E(-a', v^{2})  \;
u \;  O (+a, u) \;  O(-a, v\;)\; ( u^{2} + v^{2})  \neq  0 \; ,
\nonumber
\end{eqnarray}

\underline{in spinor space} $\qquad \qquad u_{a'++\; ,\;a--} =$
\begin{eqnarray}
=
\int_{-\infty} ^{+\infty} dv  \int _{-\infty} ^{+\infty} du
\; E (+a';, u^{2}) \; E(-a', v^{2})
\; u \;    O (+a, u) \;  O(-a, v\;) \;
 ( u^{2} + v^{2})  \equiv   0  \; .
\nonumber
\end{eqnarray}

\underline{In vector space:} $\qquad \qquad v_{a'++\; ,\;a+-} =$
\begin{eqnarray}
=
\int_{-\infty} ^{+\infty} dv  \int _{-\infty} ^{+\infty} du
\; E (+a';, u^{2}) \; E(-a', v^{2})
\; v \;    E (+a, u) \;  O(-a, v\;) \;
 ( u^{2} + v^{2})       \neq 0 \; .
\nonumber
\end{eqnarray}

\underline{in spinor space:} $\qquad \qquad v_{a'+-\; ,\;a+-} =$
\begin{eqnarray}
=
\int_{-\infty} ^{+\infty} dv  \int _{-\infty} ^{+\infty} du
\; E (+a';, u^{2}) \; O(-a', v)
\; v \;    E (+a, u^{2}) \;  O(-a, v\;) \;
 ( u^{2} + v^{2})      \equiv  0 \; .
\nonumber
\end{eqnarray}

%8
\section{Schr\"{o}dinger equation}

\hspace{5mm}
Analysis  given  on analytical  properties of Klein-Fock
wave solutions in vector and spinor
space models still retains its applicability with slight changes for the non-relativistic
Schr{\"{o}}dinger equation  as well:
\begin{eqnarray}
i \hbar {\partial \over \partial t} \Psi = - {\hbar^{2} \over 2m}\;
\left [\;
{\partial^{2} \over \partial z^{2}} + {1 \over  u^{2} + v^{2} }\;
({\partial^{2} \over  \partial u^{2}} +
{\partial^{2} \over  \partial v^{2}}  )\; \right ]
\Psi \; ,
\label{8.1a}
\end{eqnarray}

substitution for wave functions is the same
\begin{eqnarray}
\Psi (t,u,v,z) = e^{-i\epsilon t /\hbar} \; e^{ipz / \hbar} \;
U(u) V(v) \; ,
\nonumber
\end{eqnarray}

equation for $U(u)V(v)$ is
\begin{eqnarray}
\left [ \; {\hbar^{2} \over 2m}\;
({\partial^{2} \over  \partial u^{2}} +
{\partial^{2} \over  \partial v^{2}})\;
\; + \;
 (\epsilon -{p^{2} \over 2m}) ( u^{2} + v^{2}) ) \; \right ] U(u) V(v)  = 0 \; .
\label{8.1b}
 \end{eqnarray}

explicit form  of  $\hat{A}$ in $(u,v)$-representation is
\begin{eqnarray}
\hat{A} =
{1 \over 2} \; \left [ \;
{\hbar^{2} \over 2m} \;
({\partial^{2} \over  \partial u^{2}} \; - \;
{\partial^{2} \over  \partial v^{2}}  )\; + \; ( \; i\hbar {\partial \over \partial t} \;
+\; {\hbar^{2} \over 2m}\;
{\partial^{2} \over  \partial z^{2}}\; ) \; (u^{2} - v^{2}) \; \right ] \; ,
\label{8.2a}
\end{eqnarray}

in $(x,y)$ form it looks
\begin{eqnarray}
\hat{A} =
{\hbar^{2} \over 2m}  \; \left ( \; x  \;   ({\partial ^{2} \over \partial x^{2}}
- {\partial ^{2} \over \partial y^{2}} )
 + 2y \; {\partial ^{2} \over \partial x \partial y}  +
  {\partial \over \partial x} \; \right )
+ x
\; ( \; i\hbar {\partial \over \partial t} \;
+\; {\hbar^{2} \over 2m}\;
{\partial^{2} \over  \partial z^{2}}\; ) \; .
\label{8.2b}
\end{eqnarray}

%\newpage

%A
\section{
%Supplement  A.
Parametrization of  spacial spinors by
   parabolic cylindrical coordinates}

\hspace{5mm}
In  [35,37] concepts of two sorts of spatial spinors,
depending on P-orientation of primary vector space model, vector $E_{3}$
and pseudovector $\Pi_{3}$, correspond  $\eta$-spinor and $ \xi$-spinor respectively.
Procedure of extending the space model is realized simpler on the base
of curvilinear  coordinate systems.
Here let us consider this procedure in cylindrical parabolic
coordinates. In vector model they are introduced  by relations:
\begin{eqnarray}
x = {u^{2} - v^{2} \over 2 } \;\; ,\;\;
y =  u\; v \; ,  \; z = z\; ,
\nonumber
\\
v \in  [\; 0, + \infty\; ) \;\; , \;\; u,\; z \in
 (\; - \infty , \; + \infty \;  ) \; ;
\label{A.1}
\end{eqnarray}

\noindent with graphical illustration in  Fig. A1.

\vspace{3mm}
\unitlength=0.5mm
\begin{picture}(120,70)(-60,0)
\special{em:linewidth 0.4pt}
\linethickness{0.4pt}

\put(+10,+30){\vector(+1,0){100}}
\put(+110,+25){$u$}  \put(+60,0){\vector(0,+1){70}}
\put(+62,+70){$v$} \put(+20,+30){\line(0,+1){30}}
\put(+30,+30){\line(0,+1){30}} \put(+40,+30){\line(0,+1){30}}
\put(+50,+30){\line(0,+1){30}} \put(+70,+30){\line(0,+1){30}}
\put(+80,+30){\line(0,+1){30}} \put(+90,+30){\line(0,+1){30}}
\put(+100,+30){\line(0,+1){30}}
\end{picture}

\vspace{-5mm}
\begin{center}
{\bf Fig. A1.}
\end{center}

\vspace{5mm}

\noindent
It suffices to use the semi-plane shown in Fig. 1 to cove the whole plane $(x,y)$

Spacial spinor $\xi $ is given in $(u, v, z)$-coordinates as
\begin{eqnarray}
\xi (u,v,z)  = \left | \begin{array}{c}
\sqrt{ \sqrt{z^{2} + (u^{2} + v^{2})^{2} / 4} \;+
z }  \;\; \; e^{-i\gamma /2}  \\[3mm]
\sqrt{\sqrt{z^{2} + (u^{2} + v^{2})^{2} / 4 }\; -
z } \;\; \;  e^{+i\gamma /2} \end{array} \right | \; , \;\;
e^{i\gamma /2} = {u + i v \over \sqrt{ u^{2} + v^{2} }} \; ;
\label{A.2a}
\end{eqnarray}

\noindent
here the factor $e^{i\gamma /2}$ belongs to upper complex half-plan (in the case of
spinor model, it will cover tho whole complex plane).
At the plane $z=0$ (designated by $\Pi ^{+\cap -}$),  spinor $\xi$ is given  by
\begin{eqnarray}
\xi ^{+\cap -} (u,v,z=0) = {1 \over \sqrt{2}}  \; \left | \begin{array}{c}
 u - i\;  v \\ u + i \; v  \end{array}  \right | \; .
\label{A.2b}
\end{eqnarray}

Spatial spinor of the type $\eta$  is given by
\begin{eqnarray}
\eta^{\sigma }  (u,v,z) = \left |  \begin{array}{c}
\sqrt{
\sqrt{z^{2} + (u^{2} + v^{2})^{2}/ 4} -
(u^{2} + v^{2})/2}  \;\;  (\sigma  \; e^{-i\gamma /2} ) \\[3mm]
\sqrt{ \sqrt{z^{2}  + (u^{2} + v^{2})^{2} /4} +
(u^{2} + v^{2})/2 }\;\; (\; e^{-i\gamma /2} \;)
\end{array} \right |
\label{A.3a}
\end{eqnarray}

\noindent $\sigma= +1$ corresponds to upper semi-space $(z>0)$,
$\sigma= -1$ corresponds to lower semi-space $(z<0)$.
To  the plane $(z=0)$ corresponds the  simpler spinor
\begin{eqnarray}
\eta ^{+\cap -} (u,v,z=0) = \left |  \begin{array}{c}
0 \\ u  + i \; v  \end{array} \right | \; .
\label{A.3b}
\end{eqnarray}

The way to parameterize the vector  $(x,y)$-plane by $(u,v)$-coordinates
prescribes the following identification rules for the domain  $G(u, v)$ :

\vspace{5mm}
\unitlength=0.6mm
\begin{picture}(120,70)(-40,0)
\special{em:linewidth 0.4pt}
\linethickness{0.4pt}

\put(+10,+30){\vector(+1,0){100}}
\put(+110,+25){$u$}  \put(+60,0){\vector(0,+1){70}}
\put(+62,+70){$v$} \put(+20,+30){\line(0,+1){30}}
\put(+30,+30){\line(0,+1){30}} \put(+40,+30){\line(0,+1){30}}
\put(+50,+30){\line(0,+1){30}} \put(+70,+30){\line(0,+1){30}}
\put(+80,+30){\line(0,+1){30}} \put(+90,+30){\line(0,+1){30}}
\put(+100,+30){\line(0,+1){30}} \put(+60,+30){\circle{2}}
\put(+20,+30){\circle*{2}} \put(+30,+30){\circle*{2}}
\put(+40,+30){\circle*{2}}  \put(+50,+30) {\circle*{2}}
\put(+70,+30){\circle*{2}}  \put(+80,+30) {\circle*{2}}
\put(+90,+30){\circle*{2}}  \put(+100,+30){\circle*{2}}

\put(+60,+30){\oval(20,10)[b]}   \put(+60,+30){\oval(40,20)[b]}
\put(+60,+30){\oval(60,30)[b]}   \put(+60,+30){\oval(80,40)[b]}
\end{picture}

\begin{center}
{\bf Fig. A2. Identification on the boundary of $G(u,v)$}
\end{center}

Transition to spinor space model is achieved through enlargement
of the region for $v$-variable:
\begin{eqnarray}
v \in  [\; 0 , \; + \infty  ) \;\;  \Longrightarrow  \;\;
v \in  ( - \infty  , \;  + \infty  ) \; ;
\label{A.4}
\end{eqnarray}

\noindent
at this the factor  $e^{+i\gamma /2}$ in (\ref{A.1})
will belong to  the whole  circle in tho complex plane.

%\vspace{5mm}
\unitlength=0.7mm
\begin{picture}(150,70)(-20,0)
\special{em:linewidth 0.4pt}
\linethickness{0.4pt}

\put(0,+60){$\tilde{G}(y_{1},y_{2})$}
\put(0,+30){\vector(+1,0){60}} \put(+60,+25){$y_{1}$}
\put(+30,0){\vector(0,+1){60}} \put(+32,+60){$y_{2}$}
\put(+30,+30){\oval(20,20)}  \put(+40,+30){\circle*{2}}
\put(+20,+35){\vector(0,-1){5}}  \put(+20,+25){\vector(0,+1){5}}

\put(+70,29){\line(+1,0){9}} \put(+70,31){\line(+1,0){9}}
\put(+80,+30){\line(-1,+1){5}} \put(+80,+30){\line(-1,-1){5}}

\put(+90,+60){$ e^{i\gamma /2} $}
\put(+90,+30){\vector(+1,0){60}}   \put(+120,0){\vector(0,+1){60}}
\put(+120,+30){\oval(20,20)}  \put(+130,+30){\circle*{2}}
\put(+110,+35){\vector(0,-1){5}}  \put(+110,+25){\vector(0,+1){5}}
\put(+85,+35){$\gamma =+2\pi$} \put(+85,+20){$\gamma =-2\pi$}
\end{picture}

%\vspace{-5mm}

\begin{center}
{\bf Fig. A3. Transition to a spinor model  }
\end{center}

One may specially note that the identification rule in $\tilde{G}(u,v)$
covering spinor models $\tilde{\Pi}_{3}$ and $\tilde{E}_{3}$
seems simpler than that in $G(u,v)$ for a vector models
$\Pi_{3}$ and $E_{3}$:
\begin{eqnarray}
\tilde{G}(u,v)=\left \{ \begin{array}{l}
u\in(-\infty,+\infty)\;,\\
v\in(-\infty,+\infty)\;,\\
z\in(-\infty,+\infty)\;.
 \end{array} \right.
\nonumber
\end{eqnarray}

\noindent
The domain $\tilde{G}(u,v,z)$ does not require any special identification
rules on its (infinite) boundary,  in addition to Euclidean
structure of  the $(u,v,z)$-space.

The domain $\tilde{G}(u,v,z)$ looks the same both for $\Pi_{3}$  and   $E_{3}$
spinor spaces. This means that the domain $\tilde{G}(u,v,z)$ with Euclidean topology
does not determine in full the properties of spinor models, $\Pi_{3}$  or   $E_{3}$.
Same specific distinction between $\tilde{\Pi}_{3}$  and   $\tilde{E}_{3}$ models can be seen
 if one
follows how the orientation of cylindrical surfaces $\xi^{i}, \eta^{i}$ changes
 when passing from upper to  lover half-space.

Indeed, accordingly (\ref{A.2a}) and (\ref{A.3a}), $(\xi^{1},\xi^{2})$-components are oriented as follows

\vspace{-10mm}
%\unitlength=0.75 mm
\begin{picture}(120,120)(-20,0)
\special{em:linewidth 0.4pt}
\linethickness{0.4pt}

\put(+10,+95){$\xi^{1}$}
\put(+10,+70){\vector(+1,0){40}} \put(+50,+65){$u$}
\put(+30,+50){\vector(0,+1){40}} \put(+32,+90){$v$}
\put(+50,+90){\underline{$z > 0$}}
\put(+30,+70){\oval(20,20)}   \put(+40,+70){\circle*{2}}
\put(+20,+75){\vector(0,-1){5}}  \put(+20,+65){\vector(0,+1){5}}
\put(+40,+80){$\delta = \; 2$}
\put(+40,+60){$\delta = \; 1$}

\put(+10,+40){$\xi^{1}$}
\put(+10,+20){\vector(+1,0){40}} \put(+50,+15){$u$}
\put(+30,0){\vector(0,+1){40}} \put(+32,+40){$v$}
\put(+50,+40){\underline{$z < 0$}}
\put(+30,+20){\oval(20,20)}   \put(+40,+20){\circle*{2}}
\put(+20,+25){\vector(0,-1){5}}  \put(+20,+15){\vector(0,+1){5}}
\put(+40,+30){$\delta = \; 1$}
\put(+40,+10){$\delta = \; 2$}

\put(+75,+95){$\xi^{2}$}
\put(+70,+70){\vector(+1,0){40}} \put(+110,+65){$u$}
\put(+90,+50){\vector(0,+1){40}} \put(+92,+90){$v$}
\put(+110,+90){\underline{$z > 0$}}
\put(+90,+70){\oval(20,20)}   \put(+100,+70){\circle*{2}}
\put(+80,+75){\vector(0,-1){5}}  \put(+80,+65){\vector(0,+1){5}}
\put(+120,+80){$\delta = \; 1$}
\put(+120,+60){$\delta = \; 2$}

\put(+75,+40){$\xi^{2}$}
\put(+70,+20){\vector(+1,0){40}} \put(+110,+15){$u$}
\put(+90,0){\vector(0,+1){40}}   \put(+92,+40){$v$}
\put(+110,+40){\underline{$z > 0$}}
\put(+90,+20){\oval(20,20)}   \put(+100,+20){\circle*{2}}
\put(+80,+25){\vector(0,-1){5}}  \put(+80,+15){\vector(0,+1){5}}
\put(+120,+30){$\delta = \; 1$}
\put(+120,+10){$\delta = \; 2$}

\end{picture}

\begin{center}
{\bf Fig. A4. Spacial spinor $\;\xi ^{i}(u, v, z)\;$}
\end{center}

\vspace{5mm}
Instead $\eta^{1}, \eta^{2}$ for $ \tilde{\Pi}_{3}$ model are characterized by
the schemes

%\unitlength=0.75 mm
\begin{picture}(120,110)(-20,0)
\special{em:linewidth 0.4pt}
\linethickness{0.4pt}

\put(+10,+95){$\eta^{1}$}
\put(+10,+70){\vector(+1,0){40}} \put(+50,+65){$u$}
\put(+30,+50){\vector(0,+1){40}} \put(+32,+90){$v$}
\put(+50,+90){\underline{$z > 0$}}
\put(+30,+70){\oval(20,20)}   \put(+40,+70){\circle*{2}}
\put(+20,+75){\vector(0,-1){5}}  \put(+20,+65){\vector(0,+1){5}}
\put(+40,+80){$\delta = \; 2$}
\put(+40,+60){$\delta = \; 1$}

\put(+10,+40){$\eta^{1}$}
\put(+10,+20){\vector(+1,0){40}} \put(+50,+15){$u$}
\put(+30,0){\vector(0,+1){40}} \put(+32,+40){$v$}
\put(+50,+40){\underline{$z < 0$}}
\put(+30,+20){\oval(20,20)}   \put(+20,+20){\circle*{2}}
\put(+40,+25){\vector(0,-1){5}}  \put(+40,+15){\vector(0,+1){5}}
\put(+40,+30){$\delta = \; 1$}
\put(+40,+10){$\delta = \; 2$}

\put(+70,+95){$\eta^{2}$}
\put(+70,+70){\vector(+1,0){40}} \put(+110,+65){$u$}
\put(+90,+50){\vector(0,+1){40}} \put(+92,+90){$v$}
\put(+110,+90){\underline{$z > 0$}}
\put(+90,+70){\oval(20,20)}   \put(+100,+70){\circle*{2}}
\put(+80,+75){\vector(0,-1){5}}  \put(+80,+65){\vector(0,+1){5}}
\put(+120,+80){$\delta = \; 1$}
\put(+120,+60){$\delta = \; 2$}

\put(+70,+40){$\eta^{2}$}
\put(+70,+20){\vector(+1,0){40}} \put(+110,+15){$u$}
\put(+90,0){\vector(0,+1){40}}   \put(+92,+40){$v$}
\put(+110,+40){\underline{$z > 0$}}
\put(+90,+20){\oval(20,20)}   \put(+100,+20){\circle*{2}}
\put(+80,+25){\vector(0,-1){5}}  \put(+80,+15){\vector(0,+1){5}}
\put(+120,+30){$\delta = \; 1$}
\put(+120,+10){$\delta = \; 2$}

\end{picture}
%\vspace{5mm}

\begin{center}
{\bf Fig. A5. Spacial spinor $\;\eta ^{i}(u, v, z)\;$}
\end{center}

Thus, we can conclude that the concept of spinor space should be defined by
giving

1) the form of extended region $ \tilde{G}$;

2) the way to identify its boundary points

(see applying spherical or parabolic coordinates
in the same contest [37]);

3)substantial element of the concept of
spinor space consists in  indication of orientation in $\tilde{G}$ region --
the latter is determined by explicit functions $\eta^{i}$ and  $\xi^{i}$ of
$(u,v,z)$.

\vspace{5mm}

One other  aspect of the spinor space models can be clarified with the help of the
  the derivatives of $\xi^{i}(u,v,z)$ and  $\eta^{i}(u,v,z)$ with respect to $(u,v)$
\begin{eqnarray}
{\partial  \over \partial u}  \xi ^{1} =
{\xi ^{1} \over 2} \; \left (
{\rho  \over r (r  + z) }\; u +
{ i \over \rho } \; v  \right ) \; , \qquad
{\partial  \over \partial v } \xi ^{1} =
{\xi ^{1} \over 2}  \; \left (
{\rho  \over r (r  + z) } \; v  -
{ i \over \rho } \; u \right   ) \; ,
\nonumber
\\
{\partial \over \partial u} \xi ^{2} =
{\xi ^{2}\over 2}  \; \left (
{\rho  \over r (r - z) } \; u -
{ i \over \rho } v \right  ) \;, \qquad
{\partial \over \partial v} \xi ^{2} =
{\xi ^{2} \over 2}  \; \left (
{\rho  \over r (r - z) }  v  +
{ i \over \rho } \; u  \right  )\; ,
\label{A.5}
\\
{\partial \over \partial u} \;\eta ^{1} =
{\eta ^{1} \over  2} \; ( \;
 - { u \over r}  +i\;
  {v \over \rho }  \; ) \;  , \qquad
{\partial \over \partial v}  \eta ^{1} =
{\eta ^{1} \over 2}  \;(\;
- { v \over r} - i \; {u \over \rho }  \;) \; ,
\nonumber
\\
{\partial \over \partial u }  \eta ^{2} =
{\eta ^{2} \over 2}  \; (\;
+ { u \over r}  - i \; {v \over \rho }  \;) \;\; ,
\qquad
{\partial \over \partial v} \eta ^{2} =
{\eta ^{2} \over 2}\; (\;
+ { v \over r}  +  i   \; {u \over \rho } \;  )\; .
\label{A.6}
\end{eqnarray}

\noindent
One should note that relations (\ref{A.5}) and (\ref{A.6})
are  singular only on the axis $z$.
With the help of (\ref{A.5}), (\ref{A.6}) one can readily find derivatives along directions
\begin{eqnarray}
\vec{w} = ( u, v ) \; , \qquad
\vec{\nu } = ( a, b ) \; , \;\;
\nonumber
\\
(\vec{\nu } \; \vec{w}) = ( a \; u \;+\;b \; v )
\; ,\qquad  (\vec{\nu } \times  \vec{w}) =
( a \; v \;-\; b\;  u ) \; ,
\nonumber
\\
\nabla_{\vec{\nu }} \;=\vec{\nu }\; \vec{\nabla},\qquad\;\;\nabla=
(\frac{\partial}{\partial u}\;,\;\frac{\partial}{\partial v}) \; .
\nonumber
\end{eqnarray}

\noindent For $\nabla _{\vec{\nu }}\; \xi$:
\begin{eqnarray}
\nabla _{\vec{\nu }} \;\xi ^{1} = {\xi ^{1}\over 2} \;
\left [ {\rho \over r (r + z) } (\vec{\nu }\; \vec{w}) \;+\;
{i \over \rho } (\vec{\nu } \times \vec{w})\;  \right ] \; ,
\nonumber
\end{eqnarray}
\begin{eqnarray}
\nabla_{\vec{\nu }}\; \xi ^{2} = {\xi ^{2}\over 2} \left [
\; {\rho \over r (r - z) } (\vec{\nu }\; \vec{w}) \;-\;
 {i \over \rho } (\vec{\nu } \times \vec{w})\; \right ] \; .
\label{A.7}
\end{eqnarray}

\noindent For $\nabla _{\vec{\nu }} \;\eta $:
\begin{eqnarray}
\nabla_{\vec{\nu }} \; \eta ^{1} = {\eta ^{1}\over 2} \;
\left [- \; {\vec{\nu }\; \vec{w} \over r} \;+\;
{i \over \rho } (\vec{\nu } \times  \vec{w}) \; \right ] \; ,
\; \nabla_{\vec{\nu }} \; \eta ^{2} = {\eta ^{2}\over 2} \;
\left [ \; {\vec{\nu }\; \vec{w})\over r} \;-\;
{i\over \rho } (\vec{\nu } \times \vec{w})\; \right ] \; .
\nonumber
\end{eqnarray}

These equation can be considered as  fundamental equations underlying spatial spinors,
because  solutions  of these equations
provide us with  spatial spinors.

\end{document}